\documentclass{ws-mpla}

\usepackage{amsmath,amssymb,amscd,amsbsy,amsgen,amsopn,amstext,amsxtra}
\usepackage[mathscr]{eucal}
\usepackage[super,compress]{cite}
\usepackage{graphicx}
\usepackage{color}

\newcommand{\s}{\:\!}
\newcommand{\m}{\;\!}

\begin{document}
\markboth{S. Deguchi,  \& S. Okano}
{Inhomogeneous transformations in a gauged twistor formulation of a massive particle} 

%
\catchline{}{}{}{}{}
%

\title{Inhomogeneous transformations in a gauged twistor formulation of a massive particle}

\author{SHINICHI DEGUCHI}

\address{Institute of Quantum Science, College of Science and Technology, \\
Nihon University, Chiyoda-ku, Tokyo 101-8308, Japan
\\
deguchi.shinnichi@nihon-u.ac.jp}

\author{SATOSHI OKANO${}^{}$\footnote{Corresponding author} } 

\address{Department of Liberal Arts and Basic Sciences, College of Industrial Technology, \\ 
Nihon University, Narashino, Chiba 275-8576, Japan  
\\  
okano.satoshi@nihon-u.ac.jp}
\maketitle

\begin{history}
\received{Day Month Year}
\revised{Day Month Year}
\end{history}
\begin{abstract}
In this paper, we show that the mass-shell constraints 
in the gauged twistor formulation of a massive particle 
given in [Deguchi and Okano, {\it Phys.~Rev.~D} {\bf 93}, 045016 (2016) [Erratum {\bf 93}, 089906(E) (2016)]] 
are incorporated in an action automatically by extending 
the local $U(2)$ transformation to its inhomogeneous extension denoted by $IU(2)$. 
Therefore, it turns out that all the necessary constraints are incorporated into  
an action by virtue of the local $IU(2)$ symmetry of the system.  

\end{abstract}

\ccode{PACS Nos.: 
11.10.Ef, 11.30.Ly, 11.90.+t}

\numberwithin{equation}{section}

\section{Introduction}

In the mid to late 1970's, Penrose, Perj\'{e}s, and Hughston independently studied the twistor 
description of a massive spinning particle in four-dimensional Minkowski space 
\cite{Penrose1, Penrose2, Perjes1, Perjes2, Perjes3, Perjes4, Hughston}. 
In their studies, ${n(\,\ge 2)}$ twistors 
are introduced and an inhomogeneous extension of $SU(n)$, 
which is denoted by $ISU(n)$, is found as an internal symmetry group associated with massive particles. 
(See Ref. \refcite{Mason} for recent topics concerning a twistor description of massive particles.)

In the early 2000's,  
Lagrange mechanics of a massive spinning particle formulated 
using two twistors has been studied from both classical and quantum mechanical points of view 
\cite{FedZim, BALE, AFLM, FFLM, AIL, MRT, FedLuk, AFIL, MRT2}.  
In earlier formulations, however, 
all the necessary constraints are incorporated into 
actions of a massive particle by hand with the use of appropriate Lagrange multipliers.

About 8 years ago from now, 
the present authors gave a gauged twistor formulation of a massive spinning particle 
in four dimensions \cite{DegOka}.    
In this formulation, the necessary constraints 
except mass-shell constraints are automatically 
incorporated in an action by virtue of the gauge symmetry under 
a local $U(2)$ transformation. 
In contrast, the mass-shell constraints given here 
are still incorporated into the action by hand.

The purpose of this paper is to show that the mass-shell constraints can also be 
incorporated in the action automatically by extending 
the local $U(2)$ transformation to its inhomogeneous extension denoted by $IU(2)$. 
In this way, it turns out that all the necessary constraints are incorporated in 
the action on the basis of the local $IU(2)$ symmetry of the system.

This paper is organized as follows. 
In Sec. 2, we review the gauged twistor model of a massive spinning particle 
presented in Ref. \refcite{DegOka}. 
In Sec. 3, we show that the mass-shell constraints are certainly derived 
by inhomogeneous extensions of the transformation rules of twistor and some 
associated variables. 
Section 4 is devoted to conclusions. 
In Appendix A, we derive geometric formulas for the coset space $SU(2)/U(1)$.

\section{Gauged twistor model of a massive spinning particle in four dimensions 
(A review) }

Penrose, Perj\'{e}s, and Hughston introduced two or more  
independent twistors to describe a massive particle in four-dimensional Minkowski space, $\mathbf{M}$.  
About 7 years ago from now, it was proven that 
the ${n(\,\ge 2)}$-twistor expression of four-momentum vector of a massive particle 
reduces to the two-twistor expression of it by a unitary transformation \cite{OkaDeg}. 
Taking into account this fact, we now introduce two independent twistors 
$Z_{i}^{A}=(\omega_{i}^{\alpha}, \pi_{i \dot{\alpha}})$ 
$(A=0,1,2,3; \, \alpha=0,1; \, \dot{\alpha}=\dot{0}, \dot{1} )$
distinguished by the index $i$ $(i=1, 2)$ and their dual twistors  
$\bar{Z}^{i}_{A}=(\bar{\pi}^{i}_{\alpha}, \bar{\omega}{}^{i\dot{\alpha}})$.  
Here, $\bar{\pi}^{i}_{\alpha}$ and $\bar{\omega}{}^{i\dot{\alpha}}$ denote the complex conjugates 
of $\pi_{i \dot{\alpha}}$ and $\omega_{i}^{\alpha}$, respectively: 
$\bar{\pi}^{i}_{\alpha}:=\overline{\pi_{i \dot{\alpha}}}\:\!$,  
$\;\!\bar{\omega}{}^{i\dot{\alpha}}:=\overline{\omega_{i}^{\alpha}}\:\!$.  
It is assumed that $Z_{1}^{A}$ and $Z_{2}^{A}$ are not proportional to each other, i.e.,    
$Z_{1}^{A} \neq c^{\:\!} Z_{2}^{A}$ ($c \in \Bbb{C}$), and hence  
$\bar{Z}^{1}_{A} \neq \bar{c}^{\:\!} \bar{Z}^{2}_{A}$.  
The spinors $\omega_{i}^{\alpha}$ and $\pi_{i \dot{\alpha}}$ are related by  
$\omega_{i}^{\alpha}=i z^{\alpha \dot{\alpha}} \pi_{i \dot{\alpha}}$   
with $z^{\alpha \dot{\alpha}}$ being coordinates of a point in complexified Minkowski space, $\Bbb{C} \mathbf{M}$. 
In addition to the twistors and their dual twistors, we introduce 
an inhomogeneous coordinate ${\xi \,(\in \Bbb{C})}$ of a point on the coset space $\mathcal{C}:=SU(2)/U(1)$. 
The $SU(2)$ symmetry considered here is linearly realized for the pair of twistors ${(Z_{1}^{A}, Z_{2}^{A})}$. 
(For the nonlinear realization of $SU(2)$, see Appendix A.)

Now, we recall the action for a massive spinning particle found in 
Ref.~\refcite{DegOka}:    
\begin{align}
{S} &= \int_{\tau_0}^{\tau_1} d\tau 
\bigg[\;\! 
\frac{i}{2} \Big(\bar{Z}_{A}^{i} DZ^{A}_{i} 
-Z^{A}_{i} \bar{D}\bar{Z}_{A}^{i} \Big)  
\nonumber
\\
& \quad \,
-2sa -2t \Big( b^{r} \mathcal{V}_{r}{}^{3} -\dot{\xi} e_{\xi}{}^{3} -\Dot{\Bar{\xi}} e_{\bar{\xi}}{}^{3} \Big) 
-\frac{1}{\mathsf{e}} g_{\xi\bar{\xi}} D\xi D\bar{\xi} -\frac{k^2}{2} \mathsf{e}
\nonumber
\\
& \quad \, 
+h \Big( \epsilon^{ij} \pi_{i \dot{\alpha}} \pi_{j}^{\dot{\alpha}} -\sqrt{2} \:\! m e^{i\varphi} \Big) 
+\bar{h} \Big( \epsilon_{ij} \bar{\pi}^{i}_{\alpha} \bar{\pi}^{j \alpha} -\sqrt{2} \:\! m e^{-i\varphi} \Big)
\bigg] \,,
\label{2.1}
\end{align}
with 
\begin{subequations}
\label{2.2}
\begin{align}
DZ^{A}_{i} &:=\dot{Z}^{A}_{i} -iaZ^{A}_{i} -ib_{i}{}^{j} Z^{A}_{j} \,, 
\label{2.2a}
\\
\bar{D}\bar{Z}_{A}^{i} &:=\dot{\bar{Z}}_{A}^{i} +ia\bar{Z}_{A}^{i} +i\bar{Z}_{A}^{j} b_{j}{}^{i} \,,
\label{2.2b}
\\
D\xi &:=\dot{\xi}-b^{r} K_{r}{}^{\xi} \,, 
\label{2.2c}
\\ 
D\bar{\xi} &:=\dot{\bar{\xi}}-b^{r} K_{r}{}^{\bar{\xi}} \,. 
\label{2.2d}
\end{align}
\end{subequations}
Here, $Z^{A}_{i}$, $\bar{Z}_{A}^{i}$, and $\xi$ are understood to be 
complex scalar fields on the one-dimensional parameter space 
${\mathcal{T}:=\{  \tau\;\! |\, \tau_{0} \leq \tau \leq \tau_{1} \} }$ of a particle's worldline, 
$\varphi$ is a real scalar field on $\mathcal{T}$, 
$h$ is a complex scalar-density field of weight 1 on $\mathcal{T}$, 
$\mathsf{e}$ is a positive real scalar-density field of weight 1 on $\mathcal{T}$, 
and $a$ and $b^{r}$ ${(r=1,2,3)}$ are real scalar-density fields of weight 1 on $\mathcal{T}$.  
The field $a$ is regarded as a one-dimensional $U(1)$ gauge field, 
while the fields $b^{r}$ are regarded as one-dimensional $SU(2)$ gauge fields. 
The $b_{i}{}^{j}$ $(i,j=1,2)$ are the matrix elements of the matrix 
$b:=\sum_{r=1}^{3} b^{r} \sigma_{r}$ defined using the Pauli matrices $\sigma_{r}$.  
The field $\mathsf{e}$ plays the role of an einbein field.  
The specific forms of $\mathcal{V}_{r}{}^{3}$, $e_{\xi}{}^{3} $, $e_{\bar{\xi}}{}^{3}$, 
and $g_{\xi\bar{\xi}}$ are given 
in Eqs. (\ref{A10c}), (\ref{A5a}), (\ref{A5b}), and (\ref{A6b}), respectively (see Appendix A). 
The three pairs of $K_{r}{}^{\xi}$ and $K_{r}{}^{\bar{\xi}}$ distinguished by $r$ 
constitute the $SU(2)$ Killing vectors on $\mathcal{C}$.  
In addition, $m$ is a constant mass parameter, $s$ and $t$ are real constants, 
and $k$ is a positive real constant. 
A dot over a field denotes its derivative with respect to $\tau$.

The action (\ref{2.1}) remains invariant under the reparametrization 
$\tau \rightarrow \tau^{\prime}=\tau^{\prime}(\tau)$. 
In addition, this action remains invariant under the local $U(1)$ transformation  
\begin{subequations}
\label{2.3}
\begin{align}
Z_{i}^{A} &\rightarrow Z_{i}^{\prime A} =e^{i\theta(\tau)} Z_{i}^{A} \,, 
\label{2.3a}
\\
\bar{Z}{}^{i}_{A} &\rightarrow \bar{Z}^{\prime\:\! i}_{A} =e^{-i\theta(\tau)} \bar{Z}^{i}_{A} \,, 
\label{2.3b}
\\
\xi &\rightarrow \xi^{\prime}=\xi \,,
\label{2.3c}
\\
\bar{\xi} &\rightarrow \bar{\xi}^{\prime}=\bar{\xi} \,,
\label{2.3d}
\\
h &\rightarrow h^{\prime} =e^{-2i\theta(\tau)} h \,, 
\label{2.3e}
\\
\bar{h} &\rightarrow \bar{h}^{\prime} =e^{2i\theta(\tau)} \bar{h} \,, 
\label{2.3f}
\\
\varphi &\rightarrow \varphi^{\prime} =\varphi+2\theta(\tau) \,, 
\label{2.3g}
\\
a &\rightarrow a^{\prime}=a+\dot{\theta} \,, 
\label{2.3h}
\\  
b &\rightarrow b^{\prime}=b \,, 
\label{2.3i}
\\
\mathsf{e} & \rightarrow \mathsf{e}^{\prime}=\mathsf{e} \,, 
\label{2.3j}
\end{align}
\end{subequations}
with a real transformation parameter $\theta(\tau)$ satisfying $\theta(\tau_1)=\theta(\tau_0)$.  
Furthermore, the action (\ref{2.1}) remains invariant under the local $SU(2)$ transformation  
\begin{subequations}
\label{2.4}
\begin{align} 
Z_{i}^{A} &\rightarrow Z_{i}^{\prime A} =U_{i}{}^{j}(\tau) Z^{A}_{j} \,, 
\label{2.4a}
\\
\bar{Z}{}^{i}_{A} &\rightarrow \bar{Z}{}^{\prime\:\! i}_{A} =\bar{Z}{}^{j}_{A} U^{\dagger}{}_{j}{}^{i}(\tau) \,, 
\label{2.4b}
\\
\xi &\rightarrow \xi^{\prime}=\xi^{\prime}(\xi) \,, 
\label{2.4c}
\\
\bar{\xi} &\rightarrow \bar{\xi}^{\prime}=\bar{\xi}^{\prime}(\bar{\xi}) \,, 
\label{2.4d}
\\
h &\rightarrow h^{\prime} =h \,, 
\label{2.4e}
\\
\bar{h} &\rightarrow \bar{h}^{\prime} =\bar{h} \,, 
\label{2.4f}
\\
\varphi &\rightarrow \varphi^{\prime} =\varphi \,, 
\label{2.4g}
\\
a & \rightarrow a^{\prime}=a \,, 
\label{2.4h}
\\
b &\rightarrow b^{\prime}=UbU^{\dagger}-i \dot{U} U^{\dagger} , 
\label{2.4i}
\\
\mathsf{e} & \rightarrow \mathsf{e}^{\prime}=\mathsf{e} \,,  
\label{2.4j}
\end{align}
\end{subequations}
with a transformation matrix $U(\tau) \m [\s\in {SU}(2)\s]$ 
satisfying $U(\tau_1)=U(\tau_0)$.

Now, we take the unitary gauge specified by $V=1$, 
where $V$ denote coset representatives of the coset space $SU(2)/U(1)$. 
The condition $V=1$ leads to  
$\xi=\bar{\xi}=0$ (see Appendix A). 
Then, as shown in Appendix A, $\mathcal{V}_{r}{}^{s}=\delta_{r}{}^{s}$, 
$e_{\xi}{}^{3} =e_{\bar{\xi}}{}^{3} =0$, $g_{\xi \bar{\xi}}=1/2$, 
and Eq. (\ref{A13}) hold, and hence we have $D \xi=-b^{-}$ and $D\bar{\xi}=-b^{+}$ for 
$b^{\pm}:=b^{1} \mp ib^{2}$. (Here, note that $b$ can be expanded as 
$b=\sum_{r=+,-,3} b^{r} \sigma_{r}$ with 
$\sigma_{\pm}:=(\sigma_1 \pm i\sigma_2)/2$.)  
Hence, in the unitary gauge, the action (\ref{2.1}) becomes  
\begin{align}
{S} &= \int_{\tau_0}^{\tau_1} d\tau 
\bigg[\;\! 
\frac{i}{2} \Big(\bar{Z}_{A}^{i} DZ^{A}_{i} 
-Z^{A}_{i} \bar{D}\bar{Z}_{A}^{i} \Big)  
-2sa -2t b^{3} 
-\frac{1}{2\mathsf{e}} b^{\hat\imath} b^{\hat\imath} -\frac{k^2}{2} \mathsf{e}
\nonumber 
\\
& \quad \,
+h \Big( \epsilon^{ij} \pi_{i \dot{\alpha}} \pi_{j}^{\dot{\alpha}} -\sqrt{2} \:\! m e^{i\varphi} \Big) 
+\bar{h} \Big( \epsilon_{ij} \bar{\pi}^{i}_{\alpha} \bar{\pi}^{j \alpha} -\sqrt{2} \:\! m e^{-i\varphi} \Big)
\bigg] \,,
\label{2.5}
\end{align}
where 
$b^{\hat\imath} b^{\hat\imath} :=(b^1)^2 +(b^2)^2=b^{+} b^{-}$ 
($\s\hat\imath=1,2$). 
In Ref. \refcite{DegOka}, this action was referred to as 
{\em the gauged generalized Shirafuji (GGS) action}. 
In the unitary gauge, the local $SU(2)$ invariance of $S$ is spoiled, 
while $S$ remains invariant under another local $U(1)$ transformation 
\begin{subequations}
\label{2.6}
\begin{align} 
Z_{i}^{A} &\rightarrow Z_{i}^{\prime A} =\varTheta_{i}{}^{j}(\tau) Z^{A}_{j} \,, 
\label{2.6a}
\\
\bar{Z}{}^{i}_{A} &\rightarrow \bar{Z}{}^{\prime\:\! i}_{A} =\bar{Z}{}^{j}_{A} \varTheta^{\dagger}{}_{j}{}^{i}(\tau) \,, 
\label{2.6b}
\\
h &\rightarrow h^{\prime} =h \,, 
\label{2.6c}
\\
\bar{h} &\rightarrow \bar{h}^{\prime} =\bar{h} \,, 
\label{2.6d}
\\
\varphi &\rightarrow \varphi^{\prime} =\varphi \,, 
\label{2.6e}
\\
a & \rightarrow a^{\prime}=a \,, 
\label{2.6f}
\\
b &\rightarrow b^{\prime}=\varTheta b \varTheta^{\dagger} +\dot{\vartheta} \sigma_3 \,,  
\label{2.6g}
\\
\mathsf{e} & \rightarrow \mathsf{e}^{\prime}=\mathsf{e} \,,  
\label{2.6h}
\end{align}
\end{subequations}
with the matrix ${\varTheta(\tau):=\exp\{i\vartheta(\tau)\sigma_3\}}$. 
Here, $\vartheta(\tau)$ is a real transformation parameter satisfying $\vartheta(\tau_1)=\vartheta(\tau_0)$. 
The local $SU(2)$ invariance of $S$ is thus converted 
to the invariance under the local $U(1)$ transformation (\ref{2.6}) 
by means of the gauge fixing such that $V=1$. 
To avoid confusion, we hereafter refer to the transformation (\ref{2.3}) as the $U(1)_a$ transformation 
and refer to the transformation (\ref{2.6}) as the $U(1)_b$ transformation.  
Their corresponding gauge groups are simply denoted as $U(1)_a$ and $U(1)_b$.

For our purpose, it is convenient to rewrite the action (\ref{2.5}) as 
\begin{align}
{S} &= \int_{\tau_0}^{\tau_1} d\tau 
\bigg[\;\! 
\frac{i}{2} \Big( \bar{Z}_{A}^{i} \dot{Z}{}^{A}_{i} 
-Z^{A}_{i} \dot{\bar{Z}}{}_{A}^{i} \Big) 
+a \Big(\bar{Z}_{A}^{i} Z^{A}_{i} -2s \Big) 
\nonumber
\\
& \quad \,
+b^{3} \Big( \bar{Z}_{A}^{j} \sigma_{3j}{}^{k} Z^{A}_{k} -2t \Big) 
+b^{\hat{\imath}} \bar{Z}_{A}^{j} \sigma_{\hat{\imath} j}{}^{k} Z^{A}_{k} 
-\frac{1}{2\mathsf{e}} b^{\hat{\imath}} b^{\hat{\imath}} -\frac{k^2}{2} \mathsf{e}
\nonumber 
\\
& \quad \,
+h \Big(\epsilon^{ij} \pi_{i \dot{\alpha}} \pi_{j}^{\dot{\alpha}} -\sqrt{2}\:\! m e^{i\varphi} \Big) 
+\bar{h} \Big(\epsilon_{ij} \bar{\pi}^{i}_{\alpha} \bar{\pi}^{j \alpha} -\sqrt{2}\:\! m e^{-i\varphi} \Big) 
\bigg] \,. 
\label{2.7}
\end{align}
Variation of the action (\ref{2.7}) with respect to $a$, $b^3$, and $b^{\hat\imath}$ yields    
the constraints 
\begin{subequations}
\label{2.8}
\begin{align} 
&T_{0} -s =0 \,,
\label{2.8a}
\\
&T_{3} -t =0 \,, 
\label{2.8b}
\\
&T_{\hat\imath} -\frac{1}{2\mathsf{e}} b^{\hat{\imath}}=0 \,,    
\label{2.8c} 
\end{align}
\end{subequations}
respectively,   
where 
\begin{align} 
T_{0}:=\frac{1}{2} \bar{Z}_{A}^{i} Z^{A}_{i} \,, \quad 
T_{r}:=\frac{1}{2} \bar{Z}_{A}^{j} \sigma_{rj}{}^{k} Z^{A}_{k}  \;\; 
(r={\hat\imath}, 3) \,. 
\label{2.9}
\end{align}
Additionally, variation of the action (\ref{2.7}) with respect to $\mathsf{e}$ yields 
$b^{\hat\imath} b^{\hat\imath}=k^2 \mathsf{e}^2$. 
Combining this and Eq. (\ref{2.8c}), we obtain 
\begin{align}
T_{\hat\imath} T_{\hat\imath}-\frac{1}{4} k^2=0 \,. 
\label{2.10}
\end{align}
In earlier formulations  
\cite{FedZim, BALE, AFLM, FFLM, AIL, MRT, FedLuk, AFIL, MRT2}, 
constraints similar to Eqs. (\ref{2.8a}), (\ref{2.8b}), and (\ref{2.10}) have been 
incorporated into an action by hand with the help of appropriate Lagrange multipliers. 
In contrast, in our formulation, the constraints (\ref{2.8a})--(\ref{2.8c}) are   
automatically incorporated in the action $S$ on the basis of the gauge principle. 
In fact, Eq. (\ref{2.8a}) is based on the $U(1)_a$ invariance of $S$ and 
Eqs. (\ref{2.8b}) and (\ref{2.8c}) are based on the $U(1)_b$ invariance of $S$.  
For the constraint (\ref{2.10}), it is obtained in connection with the invariance of $S$ 
under the reparametrization $\tau \rightarrow \tau^\prime(\tau)$ and the ${U}(1)_b$ transformation.

Variation of $S$ with respect to $h$ and $\bar{h}$ yields the mass-shell constraints  
\begin{subequations}
\label{2.11}
\begin{align}
& \epsilon^{ij} \pi_{i \dot{\alpha}} \pi_{j}^{\dot{\alpha}} -\sqrt{2}\:\! m e^{i\varphi} =0 \,, 
\label{2.11a}
\\
& \epsilon_{ij} \bar{\pi}^{i}_{\alpha} \bar{\pi}^{j \alpha} -\sqrt{2}\:\! m e^{-i\varphi} =0 \,.  
\label{2.11b}
\end{align}
\end{subequations}
Equation (\ref{2.11b}) is the complex conjugate of Eq. (\ref{2.11a}). 
In our previous paper \cite{DegOka}, 
unlike the constraints (\ref{2.8a})--(\ref{2.8c}), 
the constraints (\ref{2.11a}) and (\ref{2.11b}) have not been considered to be that which  
are derived in relation to some gauge invariance of $S$. 
In earlier papers  
\cite{FedZim, BALE, AFLM, FFLM, AIL, MRT, FedLuk, AFIL, MRT2, DegOka}, 
the constraints (\ref{2.11a}) and (\ref{2.11b}), 
or constraints similar to them, 
have been introduced by hand without taking into account gauge symmetries.

\section{Inhomogeneous transformations} 

For convenience, we combine the transformations (\ref{2.3}) and (\ref{2.4}) into    
the following local $U(2)$ transformation: 
\begin{subequations}
\label{3.1}
\begin{align}
Z_{i}^{A} &\rightarrow Z_{i}^{\prime A} =\mathcal{U}_{i}{}^{j}(\tau) Z^{A}_{j} \,, 
\label{3.1a}
\\
\bar{Z}{}^{i}_{A} &\rightarrow \bar{Z}{}^{\prime\:\! i}_{A} =\bar{Z}{}^{j}_{A} \mathcal{U}^{\dagger}{}_{j}{}^{i}(\tau) \,.
\label{3.1b}
\\
\xi &\rightarrow \xi^{\prime}=\xi^{\prime}(\xi) \,, 
\label{3.1c}
\\
\bar{\xi} &\rightarrow \bar{\xi}^{\prime}=\bar{\xi}^{\prime}(\bar{\xi}) \,, 
\label{3.1d}
\\
h &\rightarrow h^{\prime} =e^{-2i\theta(\tau)}h \,, 
\label{3.1e}
\\
\bar{h} &\rightarrow \bar{h}^{\prime} =e^{2i\theta(\tau)}\bar{h} \,, 
\label{3.1f}
\\
\varphi &\rightarrow \varphi^{\prime} =\varphi +2\theta(\tau) \,, 
\label{3.1g}
\\
\mathsf{a} &\rightarrow \mathsf{a}^{\prime}=\mathcal{U}\mathsf{a} \, \mathcal{U}^{\dagger}
-i \:\! \dot{\mathcal{U}} \:\! \mathcal{U}^{\dagger} ,
\label{3.1h}
\\
\mathsf{e} & \rightarrow \mathsf{e}^{\prime}=\mathsf{e} \,, 
\label{3.1i}
\end{align}
\end{subequations}
where $\mathcal{U}(\tau)$ is defined by $\mathcal{U}(\tau):=e^{i\theta(\tau)} U(\tau)$, 
being an element of $U(2) \cong U(1)_a \times SU(2)$ and satisfying 
$\mathcal{U}(\tau_1)=\mathcal{U}(\tau_0)$. In addition, $\mathsf{a}$ is a one-dimensional 
$U(2)$ gauge field defined by  
$\mathsf{a}:=a\sigma_0 +b$, with $\sigma_0$ being the 2 by 2 unit matrix. 
The transformation rule (\ref{3.1h}) can be verified by using the transformation rules (\ref{2.3h}) and (\ref{2.4i}). 
As pointed out by Perj\'{e}s and Hughston independently, 
the linear transformations 
that preserve both the momentum vector  
$p_{\alpha \dot{\alpha}} :=\sum_{i=1}^{n} \bar{\pi}^{i}_{\alpha} \pi_{i \dot{\alpha}}$ $(n\geq2)$  
and the angular momentum spinor  
$\mu_{\alpha\beta} :=(i/2)\sum_{i=1}^{n} (\omega_{i\alpha} \bar{\pi}^{i}_{\beta} 
+\omega_{i\beta} \bar{\pi}^{i}_{\alpha})$ 
form a Lie group by themselves \cite{Perjes1, Perjes2, Perjes3, Perjes4, Hughston}.  
This group is a group extension of $U(n)$ defined so as to include inhomogeneous transformations,  
and it is denoted by $IU(n)$.

For a massive particle, 
the $n$-twistor expression of $p_{\alpha \dot{\alpha}}$ 
given above reduces 
to the two-twistor expression 
$p_{\alpha \dot{\alpha}} =\sum_{i=1}^{2} \bar{\pi}^{i}_{\alpha} \pi_{i \dot{\alpha}}$ 
by a unitary transformation, as was proved in Ref. \refcite{OkaDeg}. 
Therefore it is sufficient to consider only the two-twistor system in which the $IU(2)$ symmetry is realized. 
Inhomogeneous extensions of Eqs. (\ref{3.1a}) and (\ref{3.1b}) are, respectively, found to be 
 \cite{Perjes1, Perjes2, Perjes3, Perjes4, Hughston}  
\begin{subequations}
\label{3.2}
\begin{align}
Z^{A}_{i} &\rightarrow Z_{i}^{\prime A}
=\mathcal{U}_{i}{}^{j}(\tau) 
\Big(Z^{A}_{j} +\varLambda(\tau) \epsilon_{jk} I^{AB} \bar{Z}{}^{k}_{B} \Big) \,, 
\label{3.2a}
\\[3pt]
\bar{Z}{}^{i}_{A} &\rightarrow \bar{Z}{}^{\prime\:\! i}_{A} 
=\Big(\bar{Z}{}^{j}_{A} 
+\bar{\varLambda}(\tau) \epsilon^{jk} I_{AB} Z^{B}_{k} \Big) 
\:\! \mathcal{U}^{\dagger}{}_{j}{}^{i}(\tau) \,,  
\label{3.2b}
\end{align}
\end{subequations} 
where $\varLambda(\tau)$ is a complex transformation parameter 
for the local inhomogeneous transformation, 
satisfying $\varLambda(\tau_1)=\varLambda(\tau_0)$.  
(In the two-twistor system, a skew-symmetric parameter $\varLambda_{ij}$ can be 
written as $\varLambda \epsilon_{ij}$.)  
In addition, $I^{AB}$ and $I_{AB}$ denote the so-called infinity twistors defined by 
\begin{align}
I^{AB}:= \left(
\begin{array}{cc}
\! \epsilon^{\alpha\beta} &\: \! 0 \\
\! 0 & \:\! 0 
\end{array}
\:\! \right) ,
\quad \quad 
I_{AB}:= \left(\,
\begin{array}{cc}
\! 0 & \:\!\! 0  \\
\! 0 & \: \epsilon^{\dot{\alpha} \dot{\beta}} 
\end{array}
\! \! \right) .
\label{3.3}
\end{align}
If $\varLambda=0$, Eqs. (\ref{3.2a}) and (\ref{3.2b}) become Eqs. (\ref{3.1a}) and (\ref{3.1b}), respectively.

Using Eqs. (\ref{3.2}) and (\ref{3.1h}), we can derive the transformation behaviors  
of $DZ^{A}_{i}$ and $\bar{D} \bar{Z}_{A}^{i}$ under the local $IU(2)$ transformation as follows: 
\begin{subequations}
\label{3.4}
\begin{align}
DZ^{A}_{i} &\rightarrow D^{\prime} Z^{\prime A}_{i} 
=\mathcal{U}_{i}{}^{j} \Big\{ DZ^{A}_{j} 
+\dot{\varLambda} \epsilon_{jk} I^{AB} \bar{Z}{}^{k}_{B} 
\nonumber 
\\  
&
\qquad \qquad \quad  
+\varLambda I^{AB} \Big(\epsilon_{jk} \Dot{\Bar{Z}}{}^{k}_{B} -i\mathsf{a}_{j}{}^{k} \epsilon_{kl} \bar{Z}^{l}_{B} \Big) \Big\} \, , 
\label{3.4a}
\\
\bar{D}\bar{Z}_{A}^{i} &\rightarrow \bar{D}^{\prime}\bar{Z}^{\prime\:\! i}_{A} 
=\Big\{ \bar{D} \bar{Z}_{A}^{j}
+\Dot{\Bar{\varLambda}} \epsilon^{jk} I_{AB} Z_{k}^{B} 
\nonumber 
\\  
&
\qquad \qquad \quad 
+\bar{\varLambda} I_{AB} \Big(\epsilon^{jk} \Dot{Z}{}^{B}_{k} +i\mathsf{a}_{k}{}^{j} \epsilon^{kl} {Z}^{B}_{l} \Big) \Big\} 
\:\! \mathcal{U}^{\dagger}{}_{j}{}^{i} .
\label{3.4b}
\end{align}
\end{subequations}
Additionally, using Eqs (\ref{3.2}) and (\ref{3.4}) together with the formulas 
\begin{align}
I_{AB} Z^{A}_{i} Z^{B}_{j}=\frac{1}{2} \epsilon^{kl} \pi_{k\dot{\alpha}} \pi{}_{l}^{\dot{\alpha}} \epsilon_{ij} \,, 
\quad 
I^{AB} \bar{Z}{}^{i}_{A} \bar{Z}{}^{j}_{B}
=\frac{1}{2} \epsilon_{kl} \bar{\pi}{}^{k}_{\alpha} \bar{\pi}{}^{l \alpha} \epsilon^{ij} , 
\label{3.5}
\end{align}
we can show that
\begin{align}
\frac{i}{2} \Big(  
\bar{Z}{}^{\prime\:\! i}_{A} 
D^{\prime} Z^{\prime A}_{i} 
-Z^{\prime A}_{i} \bar{D}^{\prime} \bar{Z}^{\prime\:\! i}_{A} \Big) 
& =\frac{i}{2} \Big( \bar{Z}_{A}^{i} DZ^{A}_{i} -Z^{A}_{i} \bar{D} \bar{Z}_{A}^{i} \Big) 
\nonumber  
\\
& \quad \, 
-\frac{i}{2} \Big(\Dot{\Bar{\varLambda}}+2ia\bar{\varLambda} \Big) 
\epsilon^{ij} \pi_{i\dot{\alpha}} \pi{}_{j}^{\dot{\alpha}} 
\nonumber
\\
& \quad \, 
+\frac{i}{2} \Big(\dot{\varLambda}-2ia\varLambda \Big) 
 \epsilon_{ij} \bar{\pi}{}^{i}_{\alpha} \bar{\pi}{}^{j \alpha} \,. 
\label{3.6}
\end{align}
Thus we see that the integrand in the first line of Eq. (\ref{2.1})   
is not invariant under the inhomogeneous transformation,   
even when it is the global transformation specified by $\Dot{\varLambda}=0$.

Now, we note that the second and third lines in the right-hand side of Eq. (\ref{3.6}), 
which we refer to as extra terms,  
are proportional to the quantity $\epsilon^{ij} \pi_{i\dot{\alpha}} \pi{}_{j}^{\dot{\alpha}}$ or 
$\epsilon_{ij} \bar{\pi}{}^{i}_{\alpha} \bar{\pi}{}^{j \alpha}$. 
Since the same quantities are included in the action $S$ in the form of 
$h\epsilon^{ij} \pi_{i\dot{\alpha}} \pi{}_{j}^{\dot{\alpha}} 
+\bar{h} \epsilon_{ij} \bar{\pi}{}^{i}_{\alpha} \bar{\pi}{}^{j \alpha}$, 
it is possible to cancel out the extra terms by carrying out an appropriate 
modification of the transformation rules (\ref{3.1e}) and (\ref{3.1f}). 
In fact, we can completely cancel out the extra terms by modifying 
Eqs. (\ref{3.1e}) and (\ref{3.1f}) to be  
\begin{subequations}
\label{3.7}
\begin{align}
h &\rightarrow h^{\prime} =e^{-2i\theta(\tau)} 
\bigg\{ h+\frac{i}{2} \Big(\Dot{\Bar{\varLambda}}(\tau)+2ia\bar{\varLambda}(\tau) \Big) \bigg\} \,, 
\label{3.7a}
\\
\bar{h} &\rightarrow \bar{h}^{\prime} =e^{2i\theta(\tau)} 
\bigg\{ \bar{h}-\frac{i}{2} \Big(\Dot{\varLambda}(\tau)-2ia\varLambda(\tau) \Big) \bigg\} \,, 
\label{3.7b} 
\end{align}
\end{subequations}
respectively. 
Here, $h$ and $\bar{h}$ are regarded as gauge fields for the inhomogeneous transformation, 
because the transformation of $h$ and $\bar{h}$ given in Eq. (\ref{3.7}) provides the terms 
that contribute to canceling out the extra terms in Eq. (\ref{3.6}).

At the same time, we modify the transformation rule (\ref{3.1g}) 
in such a way that the mass term ${-\sqrt{2}\s m\mathcal{H}}$ 
with $\mathcal{H}:=he^{i\varphi} +\bar{h}e^{-i\varphi}$ included in $S$  
remains invariant under the simultaneous transformation that is defined by   
Eq. (\ref{3.7}) and a modified version of Eq. (\ref{3.1g}). 
The square of the both sides of the invariant condition $\mathcal{H}^{\prime}=\mathcal{H}$ leads to  
$h^{\prime} (e^{i\varphi^{\prime}})^{2} -\mathcal{H} e^{i\varphi^{\prime}} +\bar{h}^{\prime}=0$. 
This can be solved to yield
\begin{align}
\varphi &\rightarrow \varphi^{\prime}
=-i \ln \! \Bigg( \frac{\mathcal{H}\pm i 
\sqrt{4|h^{\prime}|{}^{2}-\mathcal{H}^{2}}}{2h^{\prime}} \Bigg) \,. 
\label{3.8}
\end{align}
Here, $4|h^{\prime}|{}^{2} \geq \mathcal{H}^{2}$ holds,  because  
$h^{\prime} e^{i\varphi^{\prime}} -\bar{h}^{\prime} e^{-i\varphi^{\prime}} 
=\pm i \big|h^{\prime} e^{i\varphi^{\prime}} -\bar{h}^{\prime} e^{-i\varphi^{\prime}} \big|$ 
is satisfied. Equation (\ref{3.8}) is precisely a modified version of Eq. (\ref{3.1g}).
When $\varLambda=0$, Eqs. (\ref{3.7a}), (\ref{3.7b}), and (\ref{3.8}) reduce to  
Eqs.  (\ref{3.1e}), (\ref{3.1f}), and (\ref{3.1g}), respectively.

The action $S$ turns out to be invariant 
under the local $IU(2)$ transformation 
\begin{subequations}
\label{3.9}
\begin{align}
Z^{A}_{i} &\rightarrow Z_{i}^{\prime A}
=\mathcal{U}_{i}{}^{j}(\tau) 
\Big(Z^{A}_{j} +\varLambda(\tau) \epsilon_{jk} I^{AB} \bar{Z}{}^{k}_{B} \Big) \,, 
\label{3.9a}
\\[3pt]
\bar{Z}{}^{i}_{A} &\rightarrow \bar{Z}{}^{\prime\:\! i}_{A} 
=\Big(\bar{Z}{}^{j}_{A} 
+\bar{\varLambda}(\tau) \epsilon^{jk} I_{AB} Z^{B}_{k} \Big) 
\:\! \mathcal{U}^{\dagger}{}_{j}{}^{i}(\tau) \,,  
\label{3.9b}
\\[3pt]
\xi &\rightarrow \xi^{\prime}=\xi^{\prime}(\xi) \,, 
\label{3.9c}
\\[2pt]
\bar{\xi} &\rightarrow \bar{\xi}^{\prime}=\bar{\xi}^{\prime}(\bar{\xi}) \,, 
\label{3.9d}
\\
h &\rightarrow h^{\prime} =e^{-2i\theta(\tau)} 
\bigg\{ h+\frac{i}{2} \Big(\Dot{\Bar{\varLambda}}(\tau)+2ia\bar{\varLambda}(\tau) \Big) \bigg\} \,, 
\label{3.9e}
\\
\bar{h} &\rightarrow \bar{h}^{\prime} =e^{2i\theta(\tau)} 
\bigg\{ \bar{h}-\frac{i}{2} \Big(\Dot{\varLambda}(\tau)-2ia\varLambda(\tau) \Big) \bigg\} \,, 
\label{3.9f} 
\\
\varphi &\rightarrow \varphi^{\prime}
=-i \ln \! \Bigg( \frac{\mathcal{H}\pm i 
\sqrt{4|h^{\prime}|{}^{2}-\mathcal{H}^{2}}}{2h^{\prime}} \Bigg) \,,  
\label{3.9g}
\\
\mathsf{a} &\rightarrow \mathsf{a}^{\prime}=\mathcal{U}\mathsf{a} \, \mathcal{U}^{\dagger}
-i \:\! \dot{\mathcal{U}} \:\! \mathcal{U}^{\dagger} ,
\label{3.9h}
\\
\mathsf{e} & \rightarrow \mathsf{e}^{\prime}=\mathsf{e} \,, 
\label{3.9i}
\end{align}
\end{subequations}
rather than under the local $U(2)$ transformation in Eq. (\ref{3.1}). 
In this way, the local $U(2)$ symmetry of the present twistor model 
can be extended to the local $IU(2)$ symmetry.  
 
\section{Conclusions} 

As we have seen, 
it is necessary for $S$ to include  
$\int_{\tau_{0}}^{\tau_{1}} d\tau \big[h\epsilon^{ij} \pi_{i\dot{\alpha}} \pi{}_{j}^{\dot{\alpha}}+
\bar{h}\epsilon_{ij} \bar{\pi}{}^{i}_{\alpha} \bar{\pi}{}^{j \alpha} \big]$ as an additive term 
in order that $S$ can remain invariant under the local $IU(2)$ transformation     
as a result of canceling out the extra terms in Eq. (\ref{3.6}). 
Here, it is essential that $h$ and $\bar{h}$ transform inhomogeneously as the gauge fields 
corresponding to the parameters $\bar{\varLambda}$ and $\varLambda$, respectively, as in Eq. (\ref{3.7}). 

The action $S$ also includes the mass term 
${\int_{\tau_{0}}^{\tau_{1}} d\tau [ -\sqrt{2} \s m\mathcal{H}^{\:\!} ]}$, 
which itself remains invariant under the local $IU(2)$ transformation. 
The sum of these two terms reads 
 \begin{align}
 S_{h} &:= \int_{\tau_{0}}^{\tau_{1}} d\tau \left[ h\epsilon^{ij} \pi_{i\dot{\alpha}} \pi{}_{j}^{\dot{\alpha}}+
\bar{h}\epsilon_{ij} \bar{\pi}{}^{i}_{\alpha} \bar{\pi}{}^{j \alpha} 
 -\sqrt{2} \s m\mathcal{H}^{\:\!} \right]
 \nonumber 
 \\
& \: = \int_{\tau_{0}}^{\tau_{1}} d\tau 
\left[\:\! h \! \left(  
\epsilon^{ij} \pi_{i \dot{\alpha}} \pi_{j}^{\dot{\alpha}} -\sqrt{2} \:\! m e^{i\varphi} \right) \!
+\bar{h} \! \left( \epsilon_{ij} \bar{\pi}^{i}_{\alpha} \bar{\pi}^{j \alpha} -\sqrt{2} \:\! m e^{-i\varphi} \right) 
\right] ,   
\label{4.1}
\end{align}
which is precisely the mass-shell term of $S$. 
The local $IU(2)$ invariance of $S$ holds with the aid of   
the one-dimensional gauge fields $a$, $b$, $h$, and $\bar{h}$. 

In our formulation, $S_{h}$ has been found on the basis of the local $IU(2)$ symmetry of the system.  
Therefore the mass-shell constraints (\ref{2.11a}) and (\ref{2.11b}) are considered to be
outcomes originated in the symmetry under the inhomogeneous transformation. 
We can say that all the constraints, including the mass-shell constraints,  
are automatically derived from the local $IU(2)$ symmetry in a self-contained way.

Canonical quantization of the twistor model governed by $S$ was actually carried out    
in Ref. \refcite{DegOka}. In the quantization procedure, the first-class constraints (\ref{2.8a}), (\ref{2.8b}) and (\ref{2.10}) 
are treated as conditions imposed on the physical state vector, 
after replacing functions in the constraints by the corresponding operators.
Another canonical quantization is performed after fixing gauge at the classical level. 
We expect that the convenient gauge-fixing for  quantization can be carried out by using 
the local $IU(2)$ symmetry.


\appendix
\section{}

In this paper and Ref. \refcite{DegOka}, 
the coset space $\mathcal{C}:=SU(2)/U(1)(\;\!\cong \mathbb{C}\mathbf{P}^{1})$ 
is introduced to nonlinearly realize 
the $SU(2)$ symmetry that is linearly realized for the pair of twistors $(Z_{1}^{A}, Z_{2}^{A})$. 
In the appendix, we briefly mention $\mathcal{C}$ and some formulas used 
in the gauged twistor formulation of a massive particle.

Let $\xi^{\:\!} (\:\!\in \Bbb{C})$ be an inhomogeneous coordinate of a point on $\mathcal{C}$  
and let ${V(\xi, \bar{\xi} \m)^{\s} [\s\in {SU}(2)\s]}$ be representative elements chosen from 
each left coset of $U(1)$ labeled by $\xi$. 
The left action of $U \s [\s\in {SU}(2)\s]$ on ${V(\xi, \bar{\xi} \m)}$ causes a nonlinear transformation 
$\xi \rightarrow \xi^{\prime}=\xi^{\prime}(\xi)$ in accordance with 
\begin{align}
V(\xi, \bar{\xi}\m) \rightarrow V(\xi^{\prime}, \bar{\xi}^{\prime}\:\!) 
=U V(\xi, \bar{\xi}{\m}) e^{ -i\vartheta \sigma_3} . 
\label{A1}
\end{align}
%
Here, $\sigma_3$ is the third component of the Pauli matrices, 
and $\vartheta$ is a real parameter for the $U(1)$ transformation generated by $\sigma_3$.
\cite{CWZ,SalStr,Nieuwenhuizen}

From $V(\xi, \bar{\xi}\m)$, we define $e_{\xi}{}^{r}$ and $e_{\bar{\xi}}{}^{r}$ ${(r=+,-,3)}$ by 
\begin{align}
e_{\xi}{}^{r} \sigma_{r}=-iV^{\dagger} \frac{\partial V}{\partial \xi} \,,  
\quad 
e_{\bar{\xi}}{}^{r} \sigma_{r}=-iV^{\dagger} \frac{\partial V}{\partial \bar{\xi}} \,.  
\label{A2}
\end{align}
Here, $\sigma_{\pm}$ are defined from the Pauli matrices $\sigma_1$ and $\sigma_2$ 
as $\sigma_{\pm}:=(\sigma_1 \pm i\sigma_2)/2$. 
Choosing $V$ to be an appropriate form 
\begin{align}
V\!\left(\xi, \bar{\xi}\m\right)=\exp \! \left[\s i \! \left(\s\bar{\zeta} \sigma_{+} +\zeta \sigma_{-} \right) \right]  
\label{A3}
\end{align}
with
\begin{align}
\zeta:=\frac{\xi}{|\xi|} \arctan|\xi| \,, 
\quad  
\bar{\zeta}:=\frac{\bar{\xi}}{|\xi|} \arctan|\xi| \,, 
\label{A4}
\end{align}
%
we obtain  
\begin{subequations}
\label{A5} 
\begin{alignat}{3} 
e_{\xi}{}^{+} &=0 \,, & \quad 
e_{\xi}{}^{-}  &=\frac{1}{1+|\xi|^2} \,, & \quad 
e_{\xi}{}^{3} &=\frac{-i \s \bar{\xi}}{2(1+|\xi|^2)} \,,  
\label{A5a}
\\
e_{\bar{\xi}}{}^{+} &=\frac{1}{1+|\xi|^2} \,, & \quad 
e_{\bar{\xi}}{}^{-}  &=0 \,, & \quad 
e_{\bar{\xi}}{}^{3} &=\frac{i \s {\xi}}{2(1+|\xi|^2)} \,. 
\label{A5b}
\end{alignat}
\end{subequations}
%
The so-called zweibeins 
${(e_{\xi}{}^{+}, e_{\bar{\xi}}{}^{+})}$ and ${(e_{\xi}{}^{-}, e_{\bar{\xi}}{}^{-})}$ 
lead to a proper metric tensor on $\mathcal{C}$: 
\begin{subequations}
\label{A6}
\begin{align}
g_{\xi \xi} & =e_{\xi}{}^{+} e_{\xi}{}^{-}= 0 \,,  
\label{A6a} 
\\
g_{\xi \bar{\xi}} &=g_{\bar{\xi} \xi} 
=\frac{1}{2} \! \left( e_{\xi}{}^{+} e_{\bar{\xi}}{}^{-} +e_{\bar{\xi}}{}^{+} e_{\xi}{}^{-}  \right) 
=\frac{1}{2(1+|\xi|^2)^2} \,,  
\label{A6b}
\\
g_{\bar{\xi} \bar{\xi}} &=e_{\bar{\xi}}{}^{+} e_{\bar{\xi}}{}^{-}=0 \,.  
\label{A6c}
\end{align}
\end{subequations}
%
The $\xi\bar{\xi}$-component of the metric tensor 
is precisely the Fubini-Study metric tensor on $\mathbb{C}\mathbf{P}^{1}$.

Now, suppose that $U$ is infinitely near the identity so that 
$U=1+i\epsilon^r \sigma_r$ ${(r=+, -, 3)}$ is valid with infinitesimal parameters $\epsilon^r$. 
Accordingly, $\xi^\prime$, $\bar{\xi}^\prime$, and $e^{-i\vartheta \sigma_3}$ can take the following form: 
$\xi^\prime=\xi+\epsilon^{r} K_{r}{}^{\xi}$, 
$\bar{\xi}^\prime=\bar{\xi}+\epsilon^{r} K_{r}{}^{\bar{\xi}}$, and 
$e^{-i\vartheta \sigma_3}=1-i\epsilon^r \varOmega_r \sigma_3$. 
Substituting these into Eq. (\ref{A1}), 
we obtain 
\begin{align}
 K_{r}{}^{\xi} \frac{\partial V}{\partial \xi} 
+ K_{r}{}^{\bar{\xi}} \frac{\partial V}{\partial \bar{\xi}} 
=i\sigma_r V-i\varOmega_r V\sigma_3 
\label{A7}
 \end{align}
after removing $\epsilon_r$. 
It turns out that $( K_{r}{}^{\xi}, K_{r}{}^{\bar{\xi}\;\!} )$, or more precisely  
$K_{r}:=K_{r}{}^{\xi} \partial/\partial \xi +K_{r}{}^{\bar{\xi}} \partial/\partial \bar{\xi}\s$, 
are the so-called $SU(2)$ Killing vectors on $\mathcal{C}$ 
and the $\varOmega_r$ are associated compensators. 
Multiplying Eq.(\ref{A7}) by $-iV^{\dagger}$ from the left and using Eq. (\ref{A2}) yield 
\begin{subequations}
\label{A8}
\begin{align}
\mathcal{V}_r{}^{\hat{\imath}} &=K_{r}{}^{\xi} e_{\xi}{}^{\hat{\imath}} 
+K_{r}{}^{\bar{\xi}} e_{\bar{\xi}}{}^{\hat{\imath}} 
\quad (\s \hat{\imath}=+,-) , 
\label{A8a}
\\
\mathcal{V}_r{}^{3} &=K_{r}{}^{\xi} e_{\xi}{}^{3} 
+K_{r}{}^{\bar{\xi}} e_{\bar{\xi}}{}^{3} +\varOmega_{r} \, , 
\label{A8b}
\end{align}
\end{subequations} 
%
where $\mathcal{V}_{r}{}^{s}$ $(s=\hat{\imath}, 3)$ are defined by 
\begin{align}
V^{\dagger} \sigma_r V=\mathcal{V}_{r}{}^{s} \sigma_s \,. 
\label{A9}
\end{align}
%
The matrix $\mathcal{V}$ is the adjoint representation of $V$. 
From Eq. (\ref{A9}), we have 
\begin{subequations}
\label{A10}
\begin{align}
\mathcal{V}_{r}{}^{+} &={\rm tr}  \big(V^{\dagger} \sigma_r V \sigma_{-} \big) \,, 
\label{A10a}
\\
\mathcal{V}_{r}{}^{-} &={\rm tr}  \big(V^{\dagger} \sigma_r V \sigma_{+} \big) \,,
\label{A10b}
\\
\mathcal{V}_{r}{}^{3} &=\frac{1}{2} {\rm tr}  \big(V^{\dagger} \sigma_r V \sigma_3 \big) \,.
\label{A10c}
\end{align}
\end{subequations}

In the unitary gauge specified by $V=1$, 
it follows that ${\mathcal{V}_{r}{}^{s}=\delta_{r}{}^{s}}$. 
When ${V=1}$, we see from Eq. (\ref{A3}) that ${\zeta=0}$, and hence we see from (\ref{A4}) that ${\xi=0}$. 
By substituting ${\xi=0}$ into Eqs. (\ref{A5a}) and (\ref{A5b}), they reduce to 
\begin{subequations}
\label{A11} 
\begin{align}
e_{\xi}{}^{+} =0 \,, \quad  
e_{\xi}{}^{-} =1 \,, \quad  
e_{\xi}{}^{3} =0 \,,  
\label{A11a}
\\
e_{\bar{\xi}}{}^{+} =1 \,, \quad
e_{\bar{\xi}}{}^{-} =0 \,,  \quad 
e_{\bar{\xi}}{}^{3} =0 \,. 
\label{A11b}
\end{align}
\end{subequations}
%
Accordingly, Eq. (\ref{A6}) becomes 
\begin{align}
g_{\xi \xi} = 0 \,, \quad 
g_{\xi \bar{\xi}} =g_{\bar{\xi} \xi} =\frac{1}{2} \,, \quad 
g_{\bar{\xi} \bar{\xi}} =0 \,.  
\label{A12}
\end{align}
%
Using Eqs. (\ref{A8a}) and (\ref{A11}) and ${\mathcal{V}_{r}{}^{s}=\delta_{r}{}^{s}}$, 
we can show that 
\begin{subequations}
\label{A13} 
\begin{align}
K_{+}{}^{\xi} =0 \,, \quad 
K_{-}{}^{\xi} =1 \,, \quad 
K_{3}{}^{\xi}=0 \,, 
\label{A13a}
\\
K_{+}{}^{\bar{\xi}} =1 \,, \quad   
K_{-}{}^{\bar{\xi}} =0 \,, \quad 
K_{3}{}^{\bar{\xi}}=0 \,. 
\label{A13b}
\end{align}
\end{subequations}
%
Similarly, using Eq. (\ref{A8b}), we can show that 
\begin{align}
\varOmega_{+}=0 \,, \quad 
\varOmega_{-}=0 \,, \quad 
\varOmega_{3}=1 \,. 
\label{A14}
\end{align}

{

}

\end{document}